%% file: paper.tex
\documentclass[]{fairmeta}
\usepackage{tcolorbox}
\usepackage{cuted}
\usepackage{float}
\usepackage{listings}
\usepackage{multirow}
\usepackage{siunitx}
\usepackage{algorithm}
\usepackage{algpseudocode}
\usepackage{booktabs}
\usepackage[table]{xcolor}
\usepackage{amsfonts}
\usepackage{dsfont}
\usepackage{enumitem}
\usepackage{lipsum} 

\title{Don't Waste It: Guiding Generative Recommenders with Structured Human Priors via Multi-Head Decoding}

\author[1,2,*]{Yunkai Zhang}
\author[1]{Qiang Zhang}
\author[1]{Feng (Ryan) Lin}
\author[1]{Ruizhong Qiu}
\author[1]{Hanchao Yu}
\author[1]{Jiayi (Jason) Liu}
\author[1]{Yinglong Xia}
\author[1]{Benyu Zhang}
\author[2]{Zeyu Zheng}
\author[3,\dagger]{Diji Yang}

\affiliation[1]{Meta AI}
\affiliation[2]{BAIR (UC Berkeley)}
\affiliation[3]{UC Santa Cruz}

\contribution[*]{Work done at Meta}
\contribution[\dagger]{Last author}

\abstract{
Optimizing recommender systems for objectives beyond accuracy, such as diversity, novelty, and personalization, is crucial for long-term user satisfaction. To this end, industrial practitioners have accumulated vast amounts of structured domain knowledge, which we term \emph{human priors} (e.g., item taxonomies, temporal patterns). This knowledge is typically applied through post-hoc adjustments during ranking or post-ranking. However, this approach remains decoupled from the core model learning, which is particularly undesirable as the industry shifts to end-to-end generative recommendation foundation models. On the other hand, many methods targeting these beyond-accuracy objectives often require architecture-specific modifications and discard these valuable human priors by learning user intent in a fully unsupervised manner.

Instead of discarding the human priors accumulated over years of practice, we introduce a backbone-agnostic framework that seamlessly integrates these human priors directly into the end-to-end training of generative recommenders. With lightweight, prior-conditioned adapter heads inspired by efficient LLM decoding strategies, our approach guides the model to disentangle user intent along human-understandable axes (e.g., interaction types, long- vs. short-term interests). We also introduce a hierarchical composition strategy for modeling complex interactions across different prior types. Extensive experiments on three large-scale datasets demonstrate that our method significantly enhances both accuracy and beyond-accuracy objectives. We also show that human priors allow the backbone model to more effectively leverage longer context lengths and larger model sizes.}

\date{\today}
\correspondence{Yunkai Zhang at \email{yunkai\_zhang@berkeley.edu} and Qiang Zhang at \email{qiangzhang@meta.com}}

\metadata[Code]{\url{https://github.com/zhykoties/Multi-Head-Recommendation-with-Human-Priors}}

\begin{document}

\maketitle
\input{Sections/Section1-Intro}
\input{Sections/Section2-Related-Work}
\input{Sections/Section3-Model}
\input{Sections/Section4-Experiments}
\section{Conclusion}
By integrating human priors accumulated over post-hoc business rules directly into generative recommenders' learning process, our proposed framework offers a principled approach to aligning recommendations with multifaceted human objectives. We view the capacity to leverage human priors not as a dependency, but as a strategic advantage tailored to the unique landscape of recommender systems. Unlike domains where human priors are scarce, the recommendation field possesses decades of market-validated expertise, resulting in many high-quality priors. Our experiments demonstrate that prior-conditioned adapter heads not only enhance accuracy, but also improve diversity, novelty, and personalization, which are key dimensions of the user experience that are often overlooked. The framework’s backbone-agnostic design and hierarchical composition strategy further enable flexible modeling of complex user intent, making it broadly applicable across scenarios.

While this work showcases the efficacy of using pre-defined priors, we view this as a fundamental but not the final step. A promising future direction is to establish a formal methodology for what constitutes a ``good'' prior, guiding the prior curation whether through human-in-the-loop or automated discovery of salient priors directly from data. Advancing the architectural fusion of these priors with more dynamic, context-aware mechanisms will also be critical. As the first attempt towards this goal, this work aims to strengthen the bridge between abstract human knowledge and the end-to-end optimization of large-scale generative recommenders.

\section{Acknowledgment}
We would like to thank Zhuokai Zhao and Lizhu Zhang for their support.

\clearpage
\newpage
\bibliographystyle{assets/plainnat}
\bibliography{paper}

\clearpage
\newpage
\beginappendix
\input{Sections/Appendix-Experiment-Details}
\input{Sections/Appendix-Additional-Results}

\end{document}

%% file: Sections/Section1-Intro.tex
\section{Introduction}
The goal of recommender systems extends beyond mere predictive accuracy. The importance of objectives such as novelty and diversity has long been recognized within the academic community \cite{adomavicius2005toward}, acknowledging that a successful system must balance relevance with discovery. Nevertheless, the metrics predominantly used to evaluate and optimize these systems have centered on accuracy and engagement \cite{ricci2021recommender}. This focus catalyzed significant algorithmic advancements, from collaborative filtering \cite{sarwar2001item, he2017neural} to deep sequential models \cite{kang2018self, zhai_actions_2024}. However, the prioritization of easily measurable signals has exposed a critical limitation, often termed the alignment problem: a model may predict the next interaction accurately, yet fail to align with the user's broader goals or well-being. This optimization imbalance has been shown to yield detrimental side effects, including polarization, addiction, and popularity bias, while discouraging the discovery of new user interests. This realization has accelerated a paradigm shift toward a human-centered approach. The critical question is evolving from ``Is this recommendation accurate?'' to ``Is this recommendation worth your time?'', which requires considering a richer set of objectives that extend beyond accuracy, such as diversity, novelty, and personalization \cite{said2025weredoingallwrong}.

To navigate these multifaceted objectives, industrial recommendation systems have accumulated a wide array of post-hoc adjustments applied during the ranking or post-ranking stage \cite{McDonald2023Impatient, Wu2025GAS}. We refer to this accumulated domain expertise as \emph{human priors}. For example, diversity is often enforced by greedily selecting candidates that maximize a combined function of relevance and entropy (defined over manually tuned categories). To favor high-value interactions (e.g., purchases over clicks), practitioners typically build separate value models for each interaction type and apply heuristic weighting schemes \cite{pei2019value, de2023systematic}. Similarly, balancing short-term engagement with long-term interests often involves temporal discounting heuristics or separate value models trained on different time horizons \cite{tang2019towards}. Furthermore, ensuring adequate personalization for minority users frequently relies on first identifying these minority users and then optimizing a separate value model, or boosting content based on demographic features \cite{lin2025fair}.

Recently, the field is trending towards the development of end-to-end (E2E) generative recommendation foundation models \cite{zhai_actions_2024, chen2024hllmenhancingsequentialrecommendations, zhou2025onerectechnicalreport}. While powerful, these models often attempt to learn user intent in an entirely unsupervised manner. Consequently, we still rely on the aforementioned post-hoc adjustments. However, these adjustments remain disconnected from the core representation learning process. As a result, the core model itself remains a black box, unaware of the crucial objectives. Additionally, to accommodate such adjustments, it is usually required to make specific changes to the model recommendations, which incurs additional cost. Alternative approaches attempt to explicitly address specific aspects, such as multi-interest networks for diversity \cite{Li2019MIND, cen2020controllable, Xie_REMI_2023} or disentanglement methods for interpretability \cite{Ma2019MacridVAE, guo2024DualVAE}. However, these methods typically require specialized architectures and their applicability in industry scenarios is still limited.

This dichotomy between complex post-hoc adjustments and unsupervised E2E models motivates a question: Instead of discarding the human priors accumulated over years of practice, can we integrate them directly into the learning process of generative recommender systems in a simple, effective, and interpretable manner? To this end, we propose a backbone-agnostic framework that seamlessly injects various human priors into the generative model training with lightweight adapters, by drawing inspiration from efficient decoding strategies in Large Language Models (LLMs)~\cite{cai2024medusasimplellminference}. Unlike post-hoc filtering or architecture-specific modifications, these adapter heads guide the sequential model to learn user representations that are naturally disentangled. This renders the model inherently controllable, explainable, and better aligned with complex, real-world objectives.

Our main contributions are summarized as follows:
\begin{itemize}
\item We generalize the concept of ``multi-interest'' to ``multi-faceted intent'' by demonstrating the framework's effectiveness across diverse human priors, including semantic, behavioral, temporal, and graph priors.
\item We propose a lightweight and backbone-agnostic framework that uses prior-conditioned adapter heads to disentangle multifaceted user intent in an end-to-end manner.
\item We introduce a hierarchical composition strategy to model complex interactions across different prior types, providing a flexible inductive bias for learning compositional representations.
\item Extensive experiments on three large-scale datasets demonstrate that our method not only improves standard accuracy metrics, but also yields significant improvements on other objectives, such as diversity, personalization, and user interest discovery.
\end{itemize}

%% file: Sections/Section2-Related-Work.tex
\section{Related Work}

We position our work at the intersection of generative recommendation, multi-interest and disentangled representation learning, and the integration of structured knowledge, motivated by the broader shift toward human-centered recommendation.

\subsection{Generative Recommenders}
Modeling the temporal dynamics of user behavior is a fundamental challenge in recommender systems. Early approaches used Recurrent Neural Networks (e.g., GRU4Rec~\cite{hidasi2015session}). The field shifted significantly by adopting the Transformer architecture, which offers superior scalability and capacity for modeling long-range dependencies. SASRec~\cite{kang2018self} established a strong baseline using self-attention for next-item prediction, leading to variants such as BERT4Rec~\cite{sun2019bert4rec} (bidirectional modeling) and S3Rec~\cite{zhou2020s3} (self-supervised learning).

Recently, the focus has shifted to large-scale foundational models. Generative Recommenders, such as HSTU~\cite{zhai_actions_2024}, frame recommendation as a sequential transduction task, demonstrating significant performance gains at scale. HLLM~\cite{chen2024hllmenhancingsequentialrecommendations} introduces a hierarchical approach by stacking two large language models (LLMs): an item LLM to capture item content and a user LLM to model user behavior. 

Despite these advances, the prevailing paradigm relies on encoding the user's history into a single, monolithic state vector. This representation bottleneck struggles to capture the heterogeneity and multi-faceted nature of user intent, often leading to suboptimal recommendations when interests conflict or evolve.

\subsection{Modeling Multi-Faceted User Intent}
To address the limitations of monolithic representations, works on multi-interest frameworks and disentangled representation learning emerged. They generally attempt to discover latent factors of user intent in an unsupervised manner and learn the preference distribution conditioned on these factors.

\emph{Multi-interest frameworks} aim to extract multiple vectors representing distinct user preferences from a single sequence. Many prominent models adopt a ``cluster-then-encode'' paradigm, relying on algorithms to partition the user history before encoding. For instance, MIND~\cite{Li2019MIND} employed dynamic routing via capsule networks to group interactions. ComiRec~\cite{cen2020controllable} extended this with a controllable aggregation framework, and REMI~\cite{Xie_REMI_2023} aimed to improve the stability of this process using regularization to prevent routing collapse.

\emph{Disentangled representation learning} focuses on separating the underlying factors of variation in user behavior, often using Variational Autoencoders (VAEs). For example, MacridVAE~\cite{Ma2019MacridVAE} sought to separate high-level intentions from low-level preferences, while DualVAE~\cite{guo2024DualVAE} learns disentangled multi-aspect representations for both users and items, and ensures a correspondence between each aspect of the user representation and the item representation.

While valuable, these unsupervised approaches share critical limitations. First, they primarily focus on disentangling topic interests (e.g., ``electronics'' vs. ``apparel''), often conflating other critical dimensions such as temporality or co-engagement structures. Second, the ``cluster-then-encode'' paradigm often relies on computationally intensive or potentially unstable discovery processes (e.g., dynamic routing, clustering). Third, the learned interest vectors often lack explicit semantic meaning. This lack of interpretability severely limits controllability, making it difficult to steer recommendations to align with business objectives, such as promoting more educational videos in order to comply with regulations.

\subsection{Integration of Human Priors and Structure}
There is growing recognition that integrating structured, human-understandable knowledge, or \emph{human priors}, can enhance model performance and interpretability (e.g., expert-defined in-domain taxonomy~\cite{Yang2024Bespoke}).

In recommender systems, human priors have traditionally been incorporated through rigid structures or post-hoc adjustments. Hierarchical models, such as HieRec~\cite{qi2021hierec}, use fixed, predefined item taxonomies to create a static interest hierarchy. While effective for taxonomy-based disentanglement, such methods cannot easily accommodate diverse, orthogonal priors (e.g., temporal dynamics) that do not fit neatly into item categories. Alternatively, industrial systems often rely on brittle post-hoc heuristic rules, which are decoupled from the core learning process.

\paragraph{Knowledge and Adaptation in LLMs.} In language models, there is significant work on enhancing models with external knowledge and structural biases. Methods like KnowBert~\cite{Peters2019KnowledgeEC} inject entity embeddings from knowledge bases to improve factual recall. Furthermore, introducing structural inductive biases, such as the Tree of Thoughts (ToT) framework~\cite{yao2023tree}, has been shown to improve reasoning abilities.

Drawing inspiration from these trends and efficient LLM adaptation techniques like Medusa~\cite{cai2024medusasimplellminference}, our work diverges from previous approaches by proposing an ``encode-then-project'' paradigm. We integrate diverse human priors directly into the end-to-end learning process using lightweight, prior-conditioned adapter heads. This bypasses the need for expensive unsupervised discovery or explicit history clustering, avoids rigid taxonomies, and yields representations that are inherently disentangled along interpretable and controllable axes.

%% file: Sections/Section3-Model.tex
\section{Model}

\subsection{Problem Formulation}
Let a user's interaction history be a sequence of items $x_{1:T}=(x_1, \dots, x_T)$, where $T$ is the context length, representing the number of item interactions in the history. The objective is to predict the user's future engagement over the next $\tau$ items, denoted as $\mathcal{Y}=\{y_{T+1}, \dots, y_{T+\tau}\}$.

First, a sequential encoder $f_\theta$ (e.g., a decoder-only transformer) is used to map the interaction history $x_{1:T}$ into a latent user state representation $\mathbf{h}_T \in \mathbb{R}^d$. Let $\mathcal{V}$ be the set of all candidate items, and each item $i \in \mathcal{V}$ is represented by an embedding $\mathbf{e}_i \in \mathbb{R}^d$. These item embeddings can either be ID-based (e.g., HSTU) or semantic-based (e.g., HLLM). The conventional approach computes a relevance score for each candidate item $i$ using the dot product between the user state and the item embedding:
\begin{align}
s(i \mid \mathbf{h}_T) = \mathbf{h}_T^\top \mathbf{e}_i.
\label{eq:sss}
\end{align}
The top-K items with the highest scores are then recommended to the user. This approach relies on a single representation $h_T$ to capture all facets of user intent, which may be suboptimal when interests are diverse, context-dependent, or evolving over time.

\subsection{Incorporating Human Priors via Conditioned Query Heads}

\begin{table}[t]
\caption{Examples of human priors supported by our framework.}
\label{tab: prior-types}
\centering
\small
\begin{tabular}{l p{0.7\linewidth}}
\toprule
\textbf{Prior Type} & \textbf{Description and Examples} \\
\midrule
Item & Semantic item attributes, such as product categories, content genres, or learned topic clusters. \\
Temporal & Evolution of user interests (e.g., short-term vs. long-term). \\
Event & The modality of the user-item interaction (e.g., \textit{click}, \textit{like}, \textit{purchase}, \textit{subscribe}). \\
Graph & Community-based item clusters derived from co-engagement or knowledge graphs. \\
User & User attributes such as demographics, subscription status, or clusters from a user co-interaction graph. \\
\bottomrule
\end{tabular}
\end{table}

Real-world user behavior is often characterized by specific factors that can be formalized as ``human priors''. These priors provide a structured and interpretable way to partition the interaction space along meaningful dimensions, such as item semantics, temporal dynamics, or interaction modalities (see Table~\ref{tab: prior-types}). To effectively incorporate these priors without modifying the backbone model $f_\theta$, we introduce a multi-head framework that employs multiple lightweight, prior-conditioned adapter heads to generate a set of specialized query embeddings, instead of relying on a single representation $\boldsymbol{h}_T$. Let $\mathcal{K}$ be the index set of the prior heads. With each head $k \in \mathcal{K}$ corresponding to a specific prior group (e.g., the ``Sports'' category), we can project the backbone's output $\mathbf{h}_T$ into different specialized query vectors $\mathbf{q}_1, \cdots \mathbf{q}_{|\mathcal{K}|}$. Inspired by the multi-head decoding structure of Medusa \cite{cai2024medusasimplellminference}, we implement the projection through a residual adapter :
\begin{equation}
\label{eq:head-proj}
\mathbf{q}_k = \mathbf{h}_T + \text{SiLU}\big(\mathbf{W}^{(k)} \mathbf{h}_T\big),
\end{equation}
where $\mathbf{W}^{(k)} \in \mathbb{R}^{d \times d}$ is a learnable transformation matrix and SiLU is the activation function~\cite{elfwing2018sigmoid}. We initialize each $\mathbf{W}^{(k)}$ with zeros, ensuring that all heads output the same representation as the original user state $\mathbf{h}_T$ at the beginning of training. As training progresses, each individual head specializes only when supported by the training signal, whereas the backbone model is shared among all prior heads. This design allows the backbone to process a user's entire interaction history, while each prior head is dedicated to modeling a specific subset of interactions.

\paragraph{Compatibility masking.} In our design, each head $k$ is restricted to retrieve only items compatible with its associated prior group, with the set of such items denoted by $\Omega_k\subseteq\mathcal{V}$, where the definition of $\Omega_k$ depends on the prior type. For example, for \emph{item-based} priors (e.g., categories), an item $i$ belongs to $\Omega_k$ if it is labeled with category $k$, and for \emph{event-based} priors, $\Omega_k$ includes items accessible through event type $k$. To enforce this specialization in inference, we define a score through the following compatibility masking: 
\begin{equation}
\label{eq:masked-score}
s_k(i|\boldsymbol{h}_T) \;=\;
\begin{cases}
\mathbf{q}_k^\top \mathbf{e}_i, & i \in \Omega_k,\\
-\infty, & i \notin \Omega_k.
\end{cases}
\end{equation}
This masking approach filters out all the incompatible items for the prior heads and ensures each head can focus exclusively on the subset of items aligned with its prior group. As a result, in contrast to the score in Eq.~\eqref{eq:sss}, the resulting score in Eq.~\eqref{eq:masked-score} is tailored to different items with their prior information, which leads to an explicit decomposition of user intent. Unlike conventional unsupervised approaches built on implicit latent factors \cite{Ma2019MacridVAE, guo2024DualVAE, Li2019MIND}, our method allows for more model interpretability as it guarantees that the learned representation $\boldsymbol{q}_k$ is identifiable. In addition, while conventional approaches suffer from the inherent uncertainty arising from the entanglement of user preferences and their underlying latent factors, our method mitigates this issue by disentangling this complexity into a set of more tractable sub-tasks, thereby enhancing the computational efficiency.

With compatibility masking, different query heads are specialized with distinct functional roles determined by the specified priors. Thus, when predicting for one prior group, our method can largely reduce the reliance on irrelevant features, which minimizes the mutual interference among these different objectives. As a result, the model can show stronger predictive capability for each prior group, and thus result in a performance improvement with the cooperation of the heads.

\subsection{Hierarchical Composition of Priors}
\label{sec: composition}
Practical recommendation settings often involve multiple, potentially interacting priors (e.g., combining item categories with temporal interests). Given $D$ distinct sets of priors $\{\mathcal{P}^{(1)},\ldots,\mathcal{P}^{(D)}\}$ with cardinalities $C^{(1)},\ldots,C^{(D)}$ respectively, a key challenge when generalizing the adapter mechanism (proposed previously in Eq.~\eqref{eq:head-proj}) is how to effectively capture the interactions between different prior sets while mitigating data sparsity for rare combinations.

We introduce a hierarchical composition strategy that organizes the adapters sequentially into a tree structure. This architecture enforces a coarse-to-fine specialization process, encouraging the model to first learn robust, shared intermediate representations at the upper levels before refining for specific prior combinations.

This design is motivated by Bayesian hierarchical modeling~\cite{Allenby2005HierarchicalBM}, which has the ``shrinkage'' effect, where group-level estimates are pulled towards a common mean as an effective form of regularization, preventing overfitting in rare prior combinations. Furthermore, this structural inductive bias mirrors recent advances in Large Language Models (LLMs), where hierarchical structures are employed to enhance reasoning, such as in Tree of Thoughts~\cite{yao2023tree}.

Starting with the base representation $\mathbf{z}^{(0)}=\mathbf{h}_T$, we recursively apply prior-specific residual adapters. At depth $d$, the representation corresponding to the path $(g_1,\ldots,g_d)$ is:
\begin{equation}
\label{eq:hier_d}
\mathbf{z}^{(d)}_{g_1,\ldots,g_d} = \mathbf{z}^{(d-1)}_{g_1,\ldots,g_{d-1}} + \mathcal{A}^{(d)}_{g_1,\ldots,g_d}\big(\mathbf{z}^{(d-1)}_{g_1,\ldots,g_{d-1}}\big).
\end{equation}
The final queries are the leaf nodes $\mathbf{q}_{g_1,\ldots,g_D} = \mathbf{z}^{(D)}_{g_1,\ldots,g_D}$. Here, $\mathcal{A}^{(d)}_{g_1,\ldots,g_d}$ denotes \emph{path-dependent adapters}, where the parameters at depth $d$ are conditioned on the entire upstream path $(g_1, \ldots, g_d)$. It is defined as:
\begin{equation}
\label{eq:adapter_detail}
\mathcal{A}^{(d)}_{g_1,\ldots,g_d}(\mathbf{z}) = \text{SiLU}\big(\mathbf{W}^{(d)}_{g_1,\ldots,g_d} \mathbf{z}\big) + \mathbf{e}_{g_{d-1}},
\end{equation}
where $\mathbf{W}^{(d)}_{g_1,\ldots,g_d} \in \mathbb{R}^{d\times d}$ are path-dependent weights. We also incorporate a learned group embedding $\mathbf{e}_{g_{d-1}} \in \mathbb{R}^d$. This embedding is shared across all branches that include $g_{d-1}$ (e.g., across all sub-trees rooted at ``short-term interest''), encouraging information sharing among related heads.

\paragraph{Hierarchical Compatibility.}
The eligible item set for a hierarchical head is defined by the intersection of the compatibilities across all involved priors:
\[
\Omega_{g_1,\ldots,g_D}
=\bigcap_{d=1}^D \Omega^{(d)}_{g_d}.
\]
An item is eligible for the head $(g_1,\ldots,g_D)$ if and only if it is compatible with \emph{all} involved priors along the path.

\subsection{Training Objective} \label{sec: train-obj}
For a specific head $k$, the set of positive examples is defined as the subset of future ground-truth items $\mathcal{Y}$ compatible with that head:
\begin{equation}
\label{eq:pos}
\mathcal{Y}_k = \{\, y \in \mathcal{Y}: y \in \Omega_k \,\}.
\end{equation}
Heads for which $\mathcal{Y}_k = \emptyset$ in a batch are excluded from the loss computation for that batch.

We optimize the parameters of each head using a unified loss framework, which can be instantiated as either a next-token prediction loss (for ID-based embeddings) or a contrastive learning loss (for semantic-based embeddings):
\begin{equation}
\label{eq:loss_k}
\mathcal{L}_{k,t} \;=\; -\mathds{1}_{y_{T+t} \in \mathcal{Y}_k}
\log \frac{\exp(s_k(y_{T+t}))}{\sum_{j\in \tilde{\Omega}_k} \exp(s_k(j))}.
\end{equation}
Here, $\mathcal{L}_{k,t}$ is the loss for head $k$ with $y_{T+t}$ as the positive item. $\tilde{\Omega}_k \subseteq \Omega_k$ is the set of items over which the softmax is computed. For next-token prediction, $\tilde{\Omega}_k$ can be all the compatible items $\Omega_k$. For contrastive learning, it is typically a subset containing the positives and some sampled negatives.

We restrict $\tilde{\Omega}_k$ to be only from $\Omega_k$. This \emph{in-group negative sampling} forces the head to discriminate among items within the same prior group. This naturally exposes the model to harder negatives (semantically or contextually similar items), leading to improved representations~\cite{robinson2021contrastive}.

To properly balance the contributions from different heads and prioritize near-future predictions, we introduce a reweighting scheme. The final loss is a sum over the forecast horizon, with each step weighted by a temporal discount factor:
\begin{equation}
\label{eq:total-loss}
\mathcal{L} \;=\; \sum_{t=1}^\tau \gamma^{t-1}\sum_{k\in\mathcal{K}} w_k \,\mathcal{L}_{k,t},
\end{equation}
where $\mathcal{L}_{k,t}$ is defined in Eq.~\eqref{eq:loss_k}. This formulation incorporates two mechanisms:
\begin{enumerate}
    \item \textbf{Frequency Balancing:} To mitigate the impact of data imbalance across heads and prevent common priors from dominating the loss, we normalize by the relative frequency of each combination of priors: $w_k^{\text{freq}} = \frac{|\mathcal{Y}_k|}{\sum_{j\in\mathcal{K}} |\mathcal{Y}_j|}$.
    \item \textbf{Temporal Discounting:} We apply a discount factor $\gamma\in(0,1]$ to prioritize near-future predictions, since predicting the very next item is often more critical than the more distant items, and the labels are also less noisy.
\end{enumerate}

\subsection{Inference and Score Fusion}
During inference, given a user state $\mathbf{h}_T$, we compute all prior-conditioned queries $\{\mathbf{q}_k\}_{k\in\mathcal{K}}$. For a candidate item $i$, we identify the set of eligible heads $\mathcal{H}(i)=\{k: i\in\Omega_k\}$. We then fuse the scores $\{s_k(i)\}_{k\in\mathcal{H}(i)}$ to obtain a final relevance score $S(i)$, which is then used to rank the candidates.

We adopt a \emph{maximum fusion} strategy: $S_{\max}(i)\;=\;\max_{k\in\mathcal{H}(i)} s_k(i)$, which allows the most relevant specialist to dominate\footnote{We explore alternative fusion strategies in Section \ref{sec: ablation-prior}.}. Not only is it computationally simple, but it also enhances interpretability by providing a clear explanation for the recommendation (e.g., ``this item was recommended because it strongly matches your short-term interest in electronics'').

\subsection{Implementation Details} 
\label{sec: implementation}
\begin{figure}[H]
    \centering
    \includegraphics[scale=0.12]{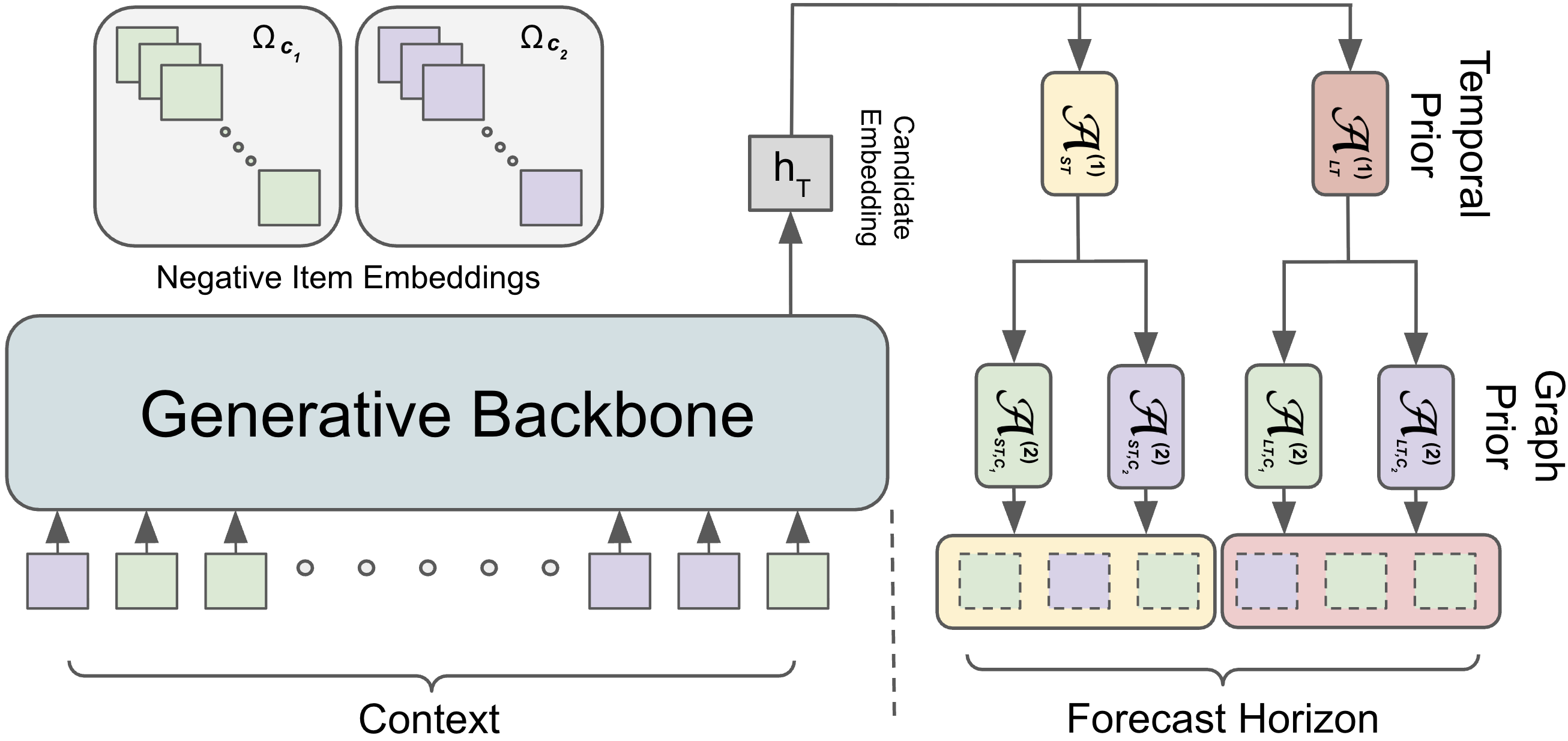}
    \caption{An instantiation of the hierarchical composition strategy with Temporal (LT/ST$_2$) and Graph Priors.}
    \label{fig: example}
\end{figure}

\textbf{A Specific Instantiation.} We illustrate a concrete instantiation of our hierarchical framework using both Temporal and Graph Priors, corresponding to the configuration used on the EB-NeRD dataset (Figure~\ref{fig: example}). For visualization, we assume two graph clusters, $C_1$ (green) and $C_2$ (purple). In this example, the forecast horizon ($\tau=6$) is divided into two equal segments (LT/ST$_{2}$), where $ST$ (short-term) is responsible for the first three targets and $LT$ (long-term) is responsible for the last three targets.

In the first layer (depth 1), the adapters specialize based on the temporal segments. The adapter $\mathcal{A}_{ST}^{(1)}$ is trained using only ground-truth items within the $ST$ segment, while $\mathcal{A}_{LT}^{(1)}$ is trained using items in the $LT$ segment. They output the intermediate representations $\mathbf{z}^{(1)}_{ST}$ and $\mathbf{z}^{(1)}_{LT}$, respectively.

In the second layer (depth 2), the specialization is further refined based on Graph Prior. For example, the adapter $\mathcal{A}_{ST,C_1}^{(2)}$ is optimized specifically for predicting items that belong to cluster $C_1$ (green) and the $ST$ segment. Similarly, $\mathcal{A}_{ST,C_2}^{(2)}$ targets $C_2$ (purple) items in the $LT$ segment. In other words, each leaf head is responsible only for the intersection of its associated priors.

Crucially, the backbone encoder processes the entire context sequence, regardless of the item's cluster. The specialization occurs only in the adapter heads. During training, we use in-group negative sampling (Section~\ref{sec: train-obj}). For the graph prior heads, negatives are sampled from the respective cluster's compatible set (e.g., $\Omega_{C_1}$ for $\mathcal{A}_{\star,C_1}^{(2)}$). The heads for Temporal Prior are compatible with all items, as any item can occur in any temporal segment. Instead, Temporal Prior restricts the training signal based on the item's position in the forecast horizon.

\textbf{Efficiency.} The proposed framework is model-agnostic and can be integrated with any generative recommender system that produces a dense user representation $\mathbf{h}_T$. The efficiency of the approach stems from the lightweight nature of the adapters\footnote{We show each head only incurs an extra $0.14\%$ of overall parameters in Section \ref{sec: main-results}.} and the parallelizability of the query computations. Formally, let $K$ denote the total number of heads and $d$ the hidden dimension. The added parameter complexity is $O(K \cdot d^2)$. The value of $K$ depends on the structure: it is the sum of categories (e.g., taxonomies), or the product for independent priors (e.g., the number of temporal segments multiplied with the number of graph clusters). During inference, adapters at the same depth level are executed in parallel, ensuring the latency overhead remains low. Per-group indices $\{\Omega_k\}$ are pre-computed and cached, enabling efficient computation of the masked scores (Eq.~\eqref{eq:masked-score}) via batched matrix multiplication.

To allow scalable distributed training on massive datasets, we implemented several critical optimizations. A significant bottleneck is the CPU memory overhead caused by replicating the dataset for each local process. We address this with a shared memory data structure, which ensures that the dataset is loaded and serialized into a contiguous byte array only once by the local master process (rank 0), which then places the data into a POSIX shared memory block. Peer processes on the same node subsequently attach to this shared memory block and access individual data objects using pre-computed byte offsets. This approach enables near zero-copy access to the data, eliminating intra-node redundancy and significantly reducing the per-node memory footprint.

Furthermore, we optimize the data loading pipeline for sequential recommendation. Existing approaches often pre-generate rolling windows, leading to substantial data redundancy. Instead, we pre-compute and store only the starting indices of the valid sample locations. It allows for more nuanced sampling strategies, such as using overlapping windows to maximize training data usage or prioritizing more recent interactions. For massive datasets exceeding available memory, we first preprocess user interaction histories and store them as individual NumPy files. During training, workers first sample a user based on predefined weights, and then access the corresponding file via NumPy memmap, which allows efficient access to file segments on disk without loading the entire file into RAM. The specific sequence window is determined according to some custom sampling strategies and only the selected window is loaded on-demand.

%% file: Sections/Section4-Experiments.tex
\section{Experiments}
\subsection{Datasets and Prior Instantiation}
To instantiate and evaluate different types of human priors (Table~\ref{tab: prior-types}), we select three real-world datasets from various domains: video (Pixel8M), e-Commerce (MerRec), and news (EB-NeRD). Detailed statistics are provided in Appendix~\ref{app:datasets} (Table~\ref{tab:dataset-stats}).

\subsubsection{Pixel8M (Video)}
Pixel8M~\cite{ChengPixel2024} is a large-scale dataset from an online video sharing platform, featuring rich multimodal item content with text and images. As the industry trend is to incorporate more modalities, this allows us to test whether our framework provides additional benefits even when the backbone model can leverage these modalities to recommend diverse contents.

\textbf{Semantic Item Prior:} To create a structured semantic prior, we consolidated the dataset's 111 highly unbalanced and sometimes redundant tags into eight high-level categories: ``Real life'', ``Informational \& Educational'', ``Fictional character'', ``Music'', ``Science \& technology'', ``Entertainment'', ``Gaming'', ``Performance \& Arts''.
To perform this task easily, consistently, and at scale, we follow the practical approach~\cite{ding2023gpt} and prompted ChatGPT to assign each original tag to one or more categories.

\subsubsection{MerRec (E-commerce)} MerRec~\cite{LiMerRec2025} is derived from the Mercari C2C marketplace. It is characterized by exceptionally long interaction sequences (there are $119756$ users who have at least $2000$ interactions) and diverse user behaviors.

\textbf{Event Prior:} Users interact with items with one of six event types: ``item view'', ``item like'', ``add to cart'', ``offer make'', ``Buy start'', and ``Buy complete''. These events represent different levels of user intent. A key challenge here is that ``offer make'', ``Buy start'', and ``Buy complete'' only occur less than $1\%$ of the time, but they are also most directly related to monetization. We use event types as priors to investigate the framework's ability to specialize on sparse, high-value signals.

\subsubsection{EB-NeRD (News)} EB-NeRD~\cite{KruseEBNeRD} is a news recommendation dataset with high-quality textual content. This dataset is less sparse than the previous two datasets, and the interaction patterns in news consumption often reflect the underlying community structures. 

\textbf{Graph Prior:} We construct an item co-engagement graph, where an edge exists if two items are interacted with by the same user. To discover underlying structural priors, for simplicity, we apply the off-the-shelf Leiden algorithm~\cite{Traag2018FromLT} from the \texttt{igraph} package, an established method for community detection that optimizes modularity and guarantees that the resulting communities are well connected. We control the influence of highly active users and merge very small clusters (details in Appendix~\ref{app:datasets}). The resulting item clusters are used as graph-based priors, testing the framework's ability to leverage community structures\footnote{While this algorithmic clustering is a practical choice for our implementation, its success also underscores the framework's tolerance to noisy or approximated priors, suggesting it can derive benefits even from imperfect structural information.}.

\subsubsection{General}\textbf{Temporal Prior.} In addition to domain-specific priors, we instantiate temporal priors, applicable across all datasets, to capture the evolution of user interests. Given a forecast horizon $\tau$, we divide it into $n$ contiguous segments. Each segment corresponds to a prior head (e.g., short-term vs. long-term), trained only on ground-truth items falling within that specific temporal segment.

\subsection{Experiment Setup}
We integrate our framework with two recent generative recommender architectures to demonstrate its generalizability:

\begin{itemize}[nosep]
    \item \textbf{HSTU}~\cite{zhai_actions_2024}: A scalable, Transformer-based architecture representing the state-of-the-art in \textbf{ID-based} modeling. It uses learned item ID embeddings, and is trained with a next-item prediction objective. We experiment with five sizes, from 12.42M to 1B parameters. We report based on size 3 by default.

    \item \textbf{HLLM}~\cite{chen2024hllmenhancingsequentialrecommendations}: A hierarchical LLM-based architecture representing the state-of-the-art in \textbf{semantic-based} modeling. It uses an Item LLM to derive item embeddings from text and visual content, and is trained with a contrastive learning objective for next item prediction.
\end{itemize}

We compare the performance of these backbone models against their counterparts enhanced with our prior-conditioned adapter framework. We also compare against the following baselines:

\begin{itemize}[nosep]
    \item \textbf{ComiRec}~\cite{cen2020controllable} is a representative \emph{multi-interest} network that outputs multiple embeddings as the different interests for each user, and uses a controllable aggregation framework to balance diversity and accuracy.

    \item \textbf{REMI}~\cite{Xie_REMI_2023} improves \emph{multi-interest} networks like ComiRec by introducing Interest-aware Hard Negative Mining and a Routing Regularization term to mitigate routing collapse.

    \item \textbf{DualVAE}~\cite{guo2024DualVAE} learns \emph{disentangled} multi-aspect representations for both users and items, and uses neighborhood-enhanced contrastive learning to ensure a direct correspondence between each aspect of item representations and user representations.
\end{itemize}

 Architectural details and hyperparameters are provided in Appendix~\ref{app:models}.\footnote{For a fair comparison, we implement ComiRec and REMI on top of the HSTU backbone, and we validated that they achieve better performances compared to the original dense layers. However, DualVAE becomes very unstable once we switch to deep encoders, instead of the shallow encoders in its \href{https://github.com/georgeguo-cn/DualVAE}{codebase}, so we stick with its original implementation.} We evaluate the recommendation accuracy using standard retrieval metrics: \textbf{Recall@K} and \textbf{NDCG@K} (Normalized Discounted Cumulative Gain). 


\subsection{Main Results} \label{sec: main-results}
Table~\ref{tab: main-results} summarizes the overall performance across the three datasets and two backbone architectures. The integration of human priors consistently improves both Recall and NDCG over different settings. Furthermore, combining multiple priors (e.g., Item + Temporal, Graph + Temporal) can lead to additional performance gains, demonstrating that our framework can effectively capture multi-faceted user intents. On EB-NeRD, with 8 temporal segments and 11 clusters, our method is able to scale to a total of 88 heads. Notably, our adapter heads are very lightweight. In the HSTU case, a single head only takes up $0.14\%$ of the overall model parameters.

\begin{table}[H]
\centering
\caption{Overall performance comparison. Human priors consistently lead to improvements over the backbone models (HSTU and HLLM). The backbones and baselines are highlighted in \textcolor{lightgray}{gray}. Note that HLLM and HSTU results are not directly comparable due to different context lengths used (See Appendix~\ref{app:datasets}), and all the baselines are run under the HSTU settings and should only be compared to HSTU.}
\label{tab: main-results}
\setlength{\tabcolsep}{5pt}
\begin{tabular}{@{} l l c c c c @{}}
\toprule
Dataset & Model & {Recall@5} & {Recall@10} & {NDCG@5} & {NDCG@10} \\
\midrule
\multirow{10}{*}{Pixel8M} 
  & \cellcolor{lightgray} HLLM & \cellcolor{lightgray} 0.84 & \cellcolor{lightgray} 1.37 & \cellcolor{lightgray} 1.46 & \cellcolor{lightgray} 1.42 \\
  & +Item & 0.91 & 1.48 & 1.57 & 1.54  \\
  & +LT/ST & 0.88 & 1.44 & 1.52 & 1.50 \\
  & +Both & \textbf{0.92} & \textbf{1.50} & \textbf{1.59} & \textbf{1.56} \\
\addlinespace
  & \cellcolor{lightgray} ComiRec & \cellcolor{lightgray} 1.04 & \cellcolor{lightgray} 1.70 & \cellcolor{lightgray} 1.80 & \cellcolor{lightgray} 1.77 \\
  & \cellcolor{lightgray} REMI & \cellcolor{lightgray} 1.13 & \cellcolor{lightgray} 1.81 & \cellcolor{lightgray} 1.99 & \cellcolor{lightgray} 1.92 \\
  & \cellcolor{lightgray} DualVAE & \cellcolor{lightgray} 0.95 & \cellcolor{lightgray} 1.49 & \cellcolor{lightgray} 1.67 & \cellcolor{lightgray} 1.60 \\
  & \cellcolor{lightgray} HSTU & \cellcolor{lightgray} 0.90 & \cellcolor{lightgray} 1.45 & \cellcolor{lightgray} 1.56 & \cellcolor{lightgray} 1.53 \\
  & +Item & 1.08 & 1.75 & 1.88 & 1.83 \\
  & +LT/ST & 1.15 & 1.90 & 1.98 & 1.96 \\
  & +Both & \textbf{1.23} & \textbf{2.00} & \textbf{2.12} & \textbf{2.09} \\
\midrule
\multirow{5}{*}{MerRec}
  & \cellcolor{lightgray} HLLM & \cellcolor{lightgray} 33.83 & \cellcolor{lightgray} 42.03 & \cellcolor{lightgray} 24.38 & \cellcolor{lightgray} 27.05 \\
  & +Event & \textbf{35.85} & \textbf{43.48} & \textbf{26.87} & \textbf{29.36} \\
\addlinespace
  & \cellcolor{lightgray} ComiRec & \cellcolor{lightgray} 40.74 & \cellcolor{lightgray} 49.49 & \cellcolor{lightgray} 30.07 & \cellcolor{lightgray} 33.01 \\
  & \cellcolor{lightgray} REMI & \cellcolor{lightgray} 41.46 & \cellcolor{lightgray} 49.27 & \cellcolor{lightgray} 31.55 & \cellcolor{lightgray} 33.15 \\
  & \cellcolor{lightgray} DualVAE & \cellcolor{lightgray} 38.36 & \cellcolor{lightgray} 48.20 & \cellcolor{lightgray} 27.29 & \cellcolor{lightgray} 31.03 \\
  & \cellcolor{lightgray} HSTU & \cellcolor{lightgray} 40.37 & \cellcolor{lightgray} 49.30 & \cellcolor{lightgray} 29.71 & \cellcolor{lightgray} 32.61 \\
  & +Event & \textbf{42.61} & \textbf{50.33} & \textbf{33.49} & \textbf{35.99} \\
\midrule
\multirow{11}{*}{EB-NeRD}
  & \cellcolor{lightgray} HLLM & \cellcolor{lightgray} 18.14 & \cellcolor{lightgray} 31.23 & \cellcolor{lightgray} 26.40 & \cellcolor{lightgray} 29.47 \\
  & +Graph & 21.09 & 36.05 & 32.19 & 35.05 \\
  & +LT/ST & 19.76 & 34.05 & 28.17 & 31.57 \\
  & +Both & \textbf{21.54} & \textbf{36.24} & \textbf{32.38} & \textbf{35.16} \\
\addlinespace
  & \cellcolor{lightgray} ComiRec & \cellcolor{lightgray} 21.47 & \cellcolor{lightgray} 35.71 & \cellcolor{lightgray} 32.18 & \cellcolor{lightgray} 34.75 \\
  & \cellcolor{lightgray} REMI & \cellcolor{lightgray} 21.61 & \cellcolor{lightgray} 34.79 & \cellcolor{lightgray} 33.11 & \cellcolor{lightgray} 34.88 \\
  & \cellcolor{lightgray} DualVAE & \cellcolor{lightgray} 18.54 & \cellcolor{lightgray} 31.24 & \cellcolor{lightgray} 28.28 & \cellcolor{lightgray} 30.61 \\
  & \cellcolor{lightgray} HSTU & \cellcolor{lightgray} 19.77 & \cellcolor{lightgray} 32.54 & \cellcolor{lightgray} 30.36 & \cellcolor{lightgray} 32.30 \\
  & +Graph & 20.78 & 34.48 & 31.59 & 33.84 \\
  & +LT/ST & 21.50 & 36.28 & 32.20 & 34.99 \\
  & +Both & \textbf{22.36} & \textbf{37.05} & \textbf{33.87} & \textbf{36.24} \\
\bottomrule
\end{tabular}
\end{table}
Both ComiRec and REMI underperform their counterparts when we inject human prior into HSTU. The reason is that instead of relying on purely unsupervised methods to discover latent interests as in multi-interest networks, our approach uses these explicit priors as a form of weak supervision to guide the model in learning different user intents as well as disentangled representations. As for DualVAE, we observe it to be very unstable when implemented based on the deep HSTU backbone. With its original shallow encoders, it only marginally outperforms HSTU on Pixel8M and even slightly lags behind HSTU on MerRec and EB-NeRD.

\subsection{Benefits Beyond Standard Metrics}
\subsubsection{A Better Accuracy-Diversity Trade-off}
We demonstrate that our method not only improves traditional ranking metrics such as Recall and NDCG, but also promotes recommendation diversity. To quantify this, we define an entropy-based metric in terms of the eight binary item categories on the Pixel8M dataset, $H\text{@K} = -\sum_{j=1}^{8} \left( \frac{n_j}{K} \log_2 \frac{n_j}{K} \right)$, where $n_j$ is the number of top-$K$ items for which the $j$-th binary feature is active. Intuitively, a higher entropy means that the top-$K$ recommended items are semantically more diverse.

\begin{figure}[H]
  \centering
    \includegraphics[scale=0.4]{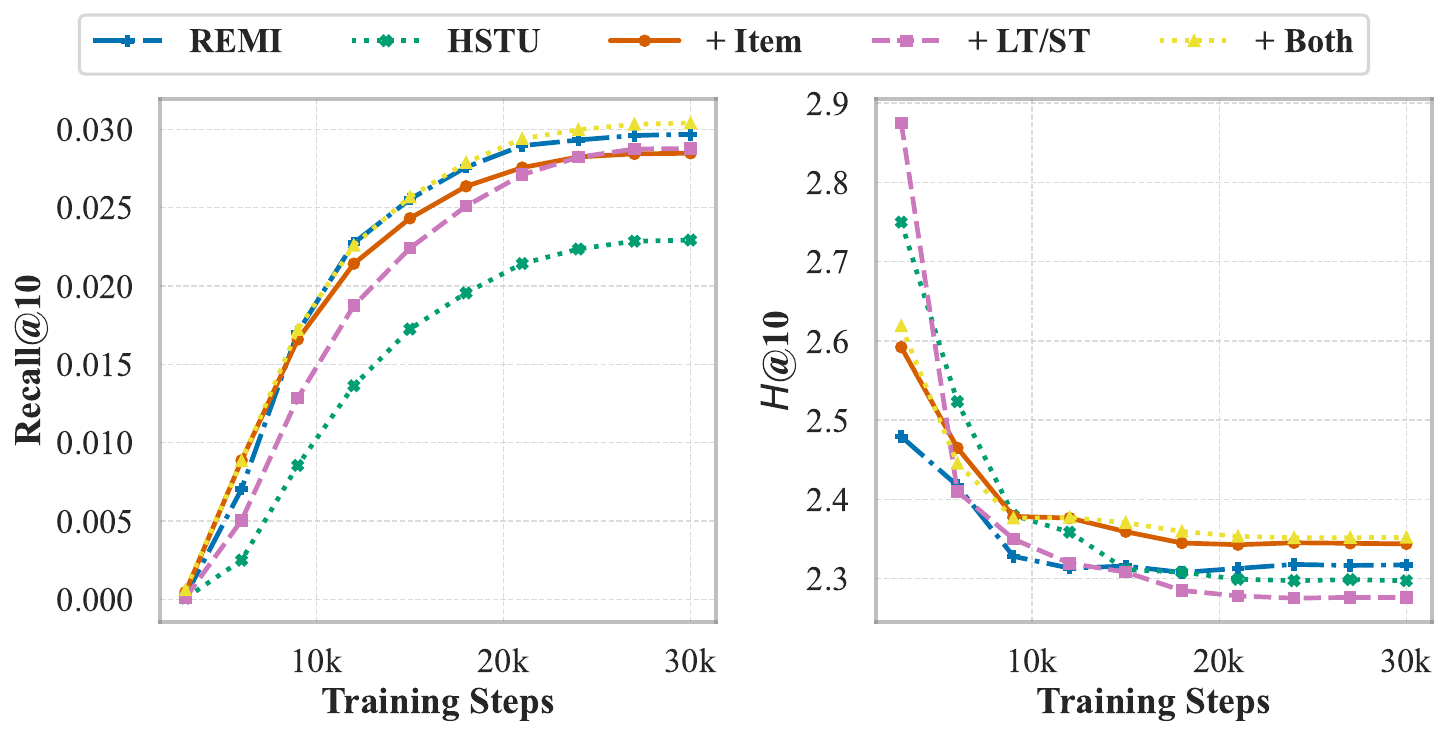}
  \caption{Evolution of entropy as training progresses on the validation set. Here HSTU is the backbone model.}
  \label{fig:exp-entropy}
\end{figure}

As shown in Figure~\ref{fig:exp-entropy}, all model variants begin training with a high entropy, which gradually declines as the training progresses. Interestingly, higher entropy typically corresponds to a lower NDCG due to the accuracy-diversity trade-off. However, we observe that injecting item priors can partially break this constraint: the variants using Item Prior and both priors simultaneously achieve a higher NDCG while maintaining a higher entropy. This balance between relevance and diversity is also observed in HLLM (Appendix Table~\ref{tab: hllm-structure-entropy}). We also include a strong multi-interest network baseline, REMI. Without explicit supervision from item priors, it only marginally improves diversity over the HSTU backbone.

\subsubsection{User Interest Exploration}

Exploration in recommender systems aims to uncover content a user may like but has not yet engaged with. Although it improves long-term user engagement, it is often believed to negatively affect near-term experience \cite{chenexplorationrs2021}, similar to the exploration versus exploitation dilemma in reinforcement learning. We argue that adding multiple heads for human priors can strike a good balance between the two. To evaluate this, we analyze the performance of our method on a targeted subset of users in the Pixel8M dataset using HLLM. Specifically, we identify users who interacted with item features during the forecast horizon of the test split, but who had never engaged with those features in their prior history. In these cases, the model's success hinges on its ability to recommend items from entirely new categories. Applying this criterion, we identified a total of 283,497 such users.

Table~\ref{tab: user-interest-exploration} compares the relative improvement over the HLLM without priors baseline achieved by different prior configurations on the standard evaluation set (\textit{All Users}) versus this subset (\textit{New Interest}). We also quantify the \textit{relative boost}, which measures how much more the method improves performance on the \textit{New Interest} subset compared to \textit{All Users}.

\begin{table}[H]
\centering
\caption{The relative improvement over the HLLM baseline for \textit{All Users} vs. \textit{New Interest}. Both priors help with user interest exploration, and Item Prior brings a even larger gain compared to LT/ST interest due to more direct supervision.}
\label{tab: user-interest-exploration}
\small
\begin{tabular}{l l r r r r}
\toprule
\textbf{Variant} & \textbf{Split} & \textbf{N@10} & \textbf{R@10} & \textbf{N@200} & \textbf{R@200} \\
\midrule

\multirow{3}{*}{LT/ST$_1$}
    & All & 7.16\% & 7.80\% & 8.10\% & 8.50\% \\
    & New & 7.41\% & 8.50\% & 8.90\% & 9.70\% \\
    & \textit{Rel. Boost} & \textit{+3.58\%} & \textit{+9.00\%} & \textit{+10.30\%} & \textit{+14.40\%} \\
\midrule

\multirow{3}{*}{LT/ST$_2$}
    & All & 5.23\% & 6.20\% & 8.00\% & 9.20\% \\
    & New & 5.37\% & 6.30\% & 8.40\% & 9.80\% \\
    & \textit{Rel. Boost} & \textit{+2.68\%} & \textit{+1.70\%} & \textit{+5.50\%} & \textit{+6.80\%} \\
\midrule

\multirow{3}{*}{LT/ST$_1$ + Item}
    & All & 8.53\% & 9.00\% & 7.90\% & 7.50\% \\
    & New & 9.87\% & 10.40\% & 9.20\% & 9.00\% \\
    & \textit{Rel. Boost} & \textbf{\textit{+15.76\%}} & \textbf{\textit{+16.50\%}} & \textbf{\textit{+17.50\%}} & \textbf{\textit{+19.50\%}} \\
\midrule

\multirow{3}{*}{LT/ST$_2$ + Item}
    & All & 9.63\% & 10.40\% & 10.70\% & 11.20\% \\
    & New & 10.22\% & 11.20\% & 11.80\% & 12.50\% \\
    & \textit{Rel. Boost} & \textit{+6.10\%} & \textit{+7.50\%} & \textit{+9.60\%} & \textit{+11.80\%} \\
\midrule

\multirow{3}{*}{LT/ST$_4$ + Item}
    & All & 8.02\% & 8.90\% & 9.80\% & 10.60\% \\
    & New & 8.49\% & 9.50\% & 10.50\% & 11.50\% \\
    & \textit{Rel. Boost} & \textit{+5.94\%} & \textit{+7.30\%} & \textit{+6.90\%} & \textit{+8.30\%} \\
\bottomrule
\end{tabular}
\end{table}

We observe that the relative improvements are consistently higher for this \textit{New Interest} subset across all configurations. Moreover, incorporating Item Prior yields substantially more benefits compared to using only Temporal Prior (LT/ST). For example, HLLM with both LT/ST$_1$ and Item Prior achieves a remarkable relative boost of +15.76\% for NDCG@10, compared to only +3.58\% for LT/ST$_1$ alone. This supports the intuition that when learning the different categories together, the user embedding will be biased against the minority items, while dedicating specific heads to the minority items allows us to retain the capacity and encourage exploration toward novel categories. Case studies in Section \ref{sec: case-study} further illustrate this insight.

\subsubsection{Personalization}
Another common issue in recommender systems is popularity bias. In some cases, it means that the behaviors of some users, whose interests are different from the majority, might receive inferior recommendations. To address this issue, we instantiated User Prior on EB-NeRD, constructed similar to Graph Prior. But here, the nodes are the users in the co-engagement graph, and an edge exists between the two users if they interacted with the same item in the train set. To avoid popular items from making the graph too dense, for each item, we random sample a maximum of 2000 users that have interacted with it before generating the edges. We then employ the Leiden algorithm \cite{Traag2018FromLT} with the modularity objective to cluster the graph into a total of 9 user groups. We define \textit{User Prior} by assigning an adapter to each group.

\begin{figure}[H]
    \centering
    \includegraphics[scale=0.3]{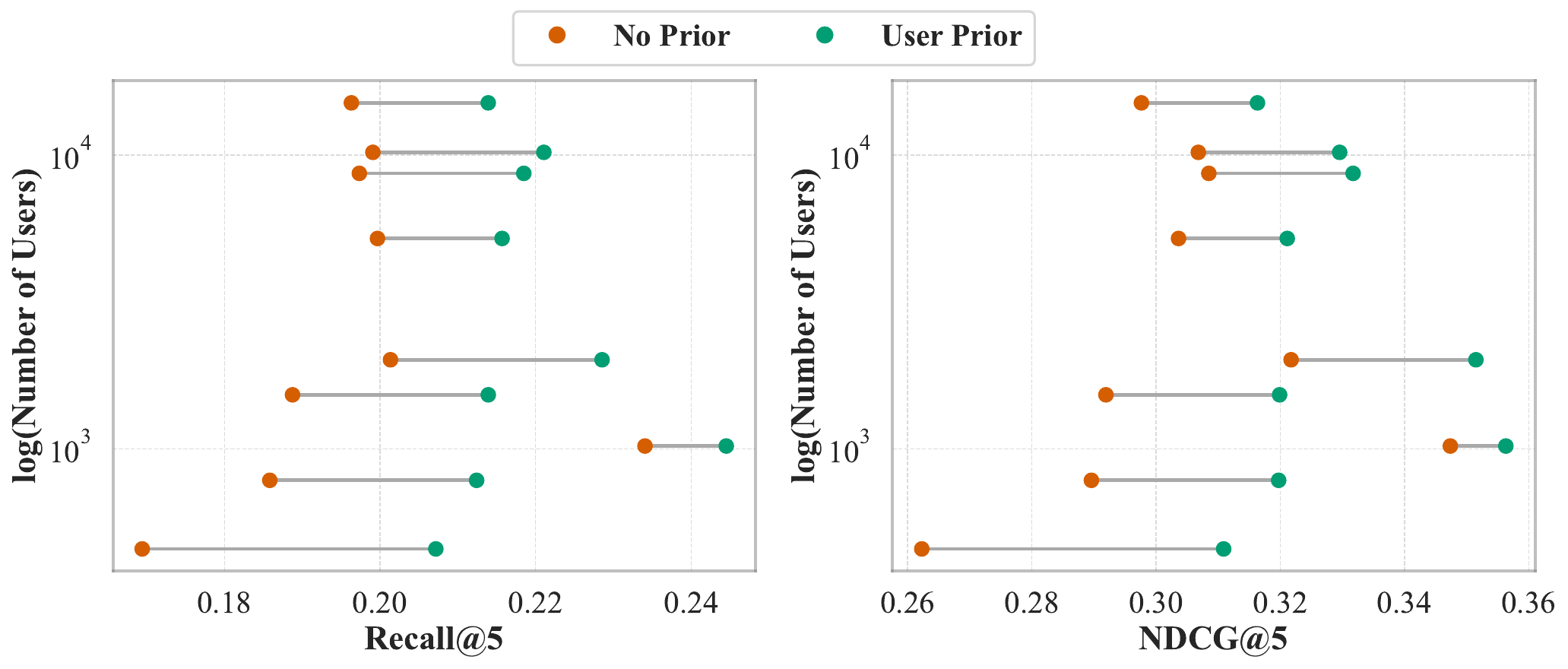}
    \caption{\textit{User Prior} leads to more personalized recommendations, especially on the minority user groups.}
    \label{fig: personalization}
\end{figure}

As shown in Figure~\ref{fig: personalization}, before introducing any prior, the minority groups with fewer users generally suffer from lower recommendation quality. However, by dedicating a separate head to each user group, we effectively lower the impact of the majority users from dominating the query representation, which allows us to better personalize the recommendations for the minority users. This results in bigger improvements for groups with fewer users, and recommendation quality looks more balanced after \textit{User Prior} is introduced.

\subsection{Scalability}

\begin{figure}[H]
    \centering
    \includegraphics[scale=0.3]{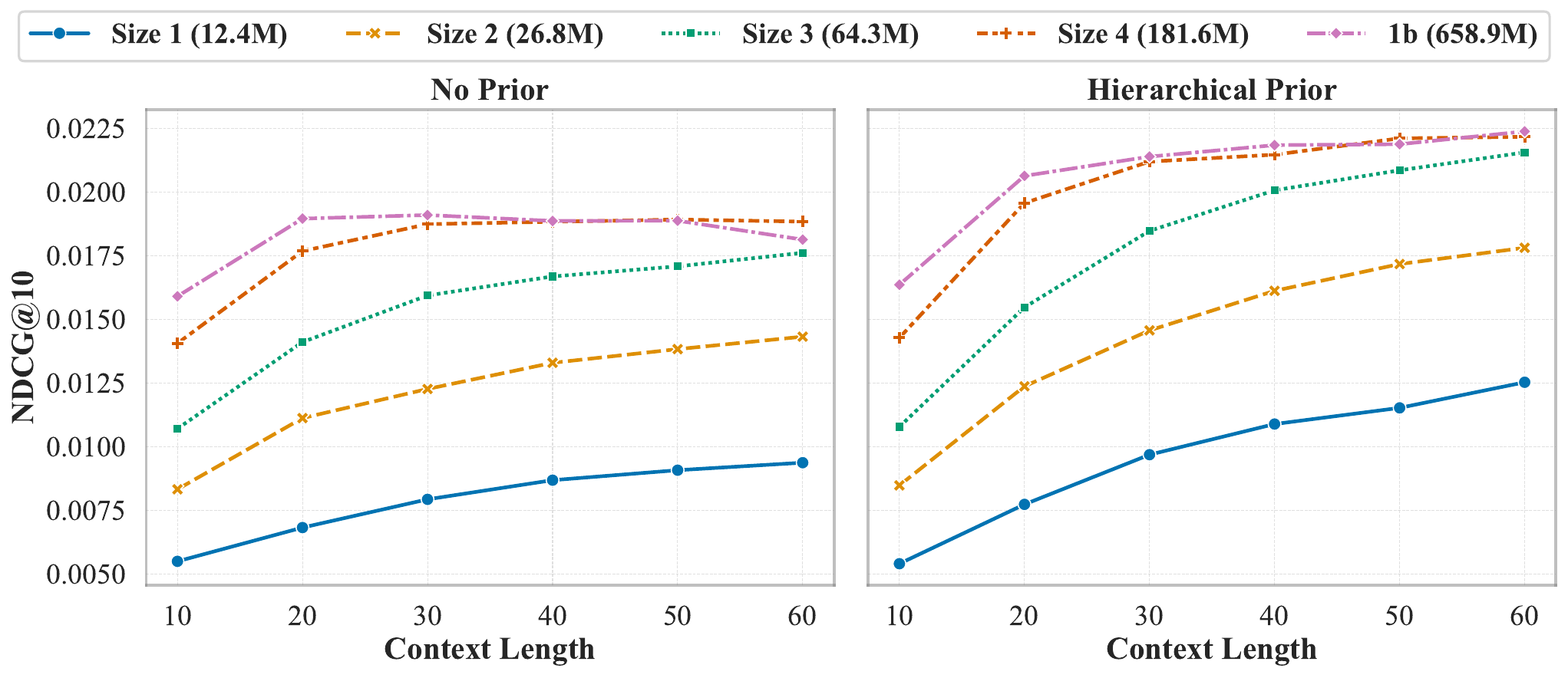}
    \caption{Scaling by context lengths and sizes for HSTU.}
    \label{fig: scale-by-context}
\end{figure}
A key trend in recommender systems is the use of longer context lengths to enhance personalization~\cite{zhang2026efficient}. While scaling laws typically require more training data to improve larger models, our experimental setup with a fixed dataset introduces a trade-off: increasing the context length reduces the number of available training windows. To investigate this, we test HSTU with longer context lengths and larger model sizes on Pixel8M. In Figure~\ref{fig: scale-by-context}, the left subplot shows that for size 4 and 1b, the base model (LT/ST$_1$, equivalent to the discounted loss) struggles to benefit from increasing context as we increase the context length beyond 20 items. However, in the right subplot, when guided by human priors, the same model architecture continues to extract performance gains from longer contexts and larger model sizes, although the magnitude of increase slowly plateaus. This finding suggests that the structural information imposed by human priors facilitates more efficient learning, allowing the model to better benefit from increasing context and larger model sizes when the amount of training data is fixed.

\subsection{Representation Space Visualization}
\label{fig:space-visual}
\begin{figure}[H]
    \centering
    \includegraphics[scale=0.3]{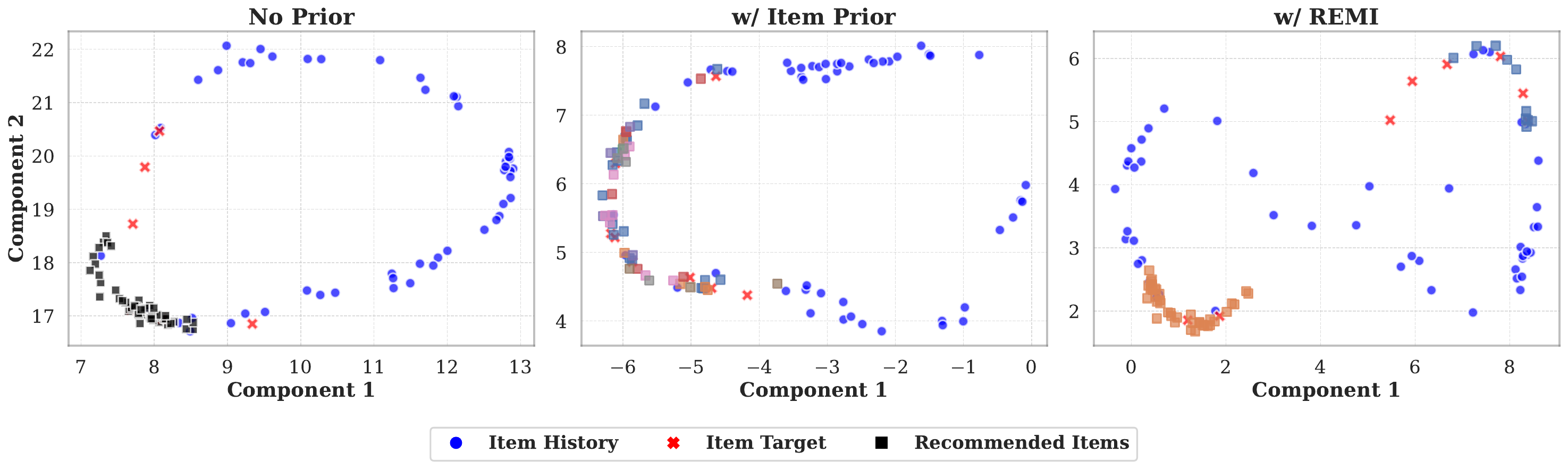}
    \caption{Visualization of the representation space for a particular user on Pixel8M. We plot item history, target items, and the top recommended items. The items recommended by different heads or interests are represented using different colors.}
    \label{fig: space-visual}
\end{figure}

Figure~\ref{fig: space-visual} presents a UMAP visualization comparing the learned representation spaces for a representative user across three models: the baseline HSTU (``No Prior''), our proposed method (``w/ Item Prior''), and REMI. For each model, we project the embeddings of the user's 50 past views, the 10 target items, and the top 50 recommended items. The UMAP is fit on the history and target embeddings. Since item embeddings are learned end-to-end with the models, the topology of the space differs for each model. For Item Prior and REMI, the recommended items are colored according to the head or interest that generates them.

We observe that the recommendations of ``No Prior'' form a single, concentrated cluster, failing to span the user's diverse targets. In contrast, ''w/ Item Prior'' leverages its different heads to generate recommendations that cover the full spectrum of the target items. For REMI, we set the number of interest to two since it achieves the best performance. However, one interest (orange squares) dominates the recommendations in the bottom-left corner with fewer target items, while the second interest (light blue squares) only partially covers the primary target cluster in the upper-right corner. This suggests that leveraging existing human priors achieves a more effective alignment between its diverse heads and the user's multifaceted interests compared to learning the multiple interests as in REMI.

\subsection{Ablations}

\subsubsection{How to Structure the Priors?}
When multiple priors are available, the strategy used to compose them significantly impacts performance. We compare the three composition strategies: Additive, Multiplicative, and Hierarchical. \emph{Additive Composition} learns the heads for each prior dimension independently. A head specializes in a single prior while remaining agnostic to others (e.g., a category-specific head optimizes for that category regardless of the time horizon). \emph{Multiplicative Composition} defines a distinct head for every element in the Cartesian product of the priors. Each head is derived independently from $\mathbf{h}_T$ using Eq.~\eqref{eq:head-proj}. We evaluate these strategies on Pixel8M, using both Item Prior and Temporal Prior. We apply these compositions on top of the HSTU backbone across five different model sizes to assess scalability and consistency.

\begin{figure}[ht]
    \centering
    \includegraphics[scale=0.3]{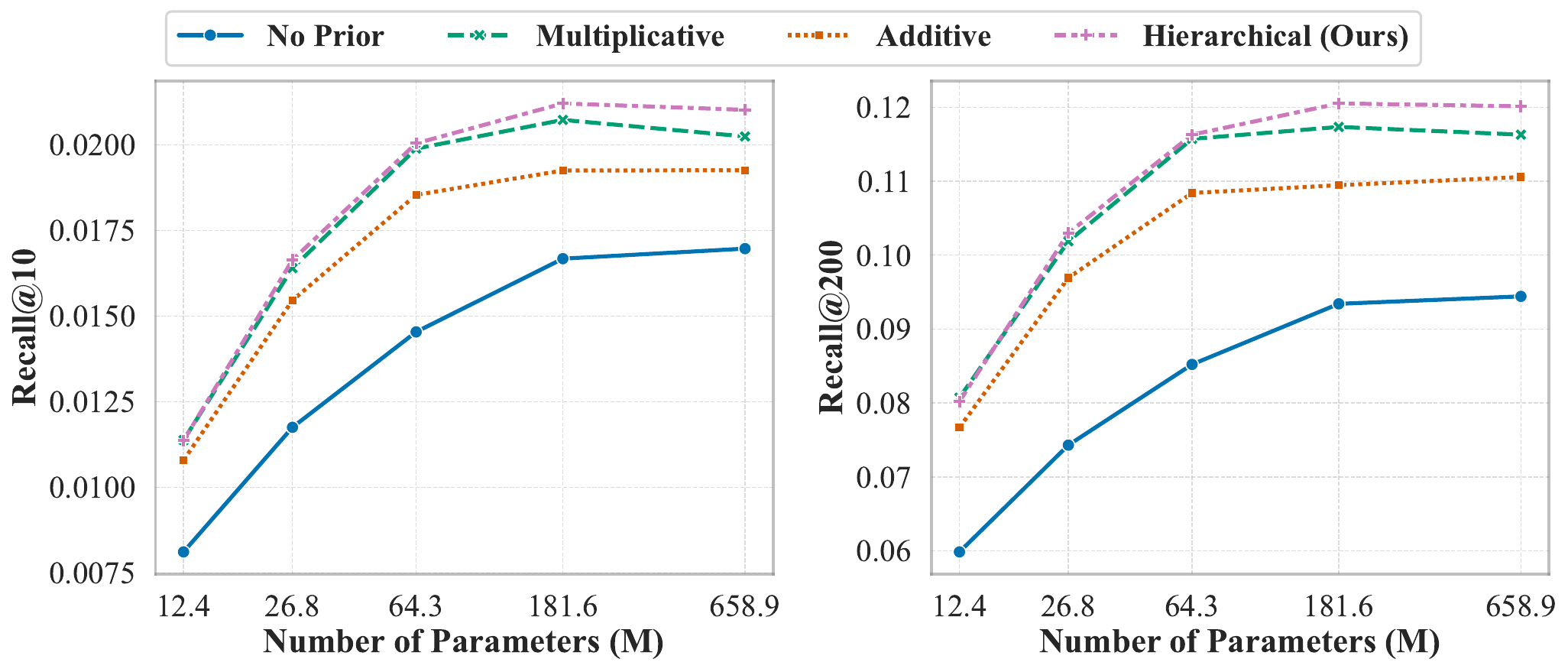}
    \caption{Comparison of composition strategies on Pixel8M across different HSTU model sizes.}
    \label{fig: how-to-structure}
\end{figure}

Figure~\ref{fig: how-to-structure} demonstrates that all three composition strategies outperform the baseline (no priors). \emph{Hierarchical Composition} consistently achieves the best results across all model sizes. Although both Hierarchical and Multiplicative have a similar parameter complexity, the Hierarchical structure provides superior performance. Specifically, Multiplicative treats the (A, B) head and the (A, C) head as completely independent entities, failing to leverage their shared ``A'' context. This can lead to data fragmentation and poor generalization, especially for rare prior combinations. This highlights the effectiveness of structural regularization and the inductive bias imposed by sequential adapters (Eq.~\eqref{eq:adapter_detail}). We observe similar trends with the HLLM backbone (Appendix Table \ref{tab: hllm-structure-entropy}).

\subsubsection{Why it works: priors or just more heads?} \label{sec: ablation-prior}
We answer two questions using Pixel8M while keeping the backbone and the number of heads fixed. First, do the gains come from human prior, or just because we benefit from more heads? We compare human priors to two variants: (i) \textbf{Random} - items are assigned to heads uniformly at random, and (ii) \textbf{All} - each item is assigned to all heads.

Second, how to fuse the scores from different heads for each item candidate at inference time? Let $s_h(i)$ be the score of item $i$ from head $h\in\mathcal{H}(i)$ (the heads that are responsible for item $i$). We consider two fusion methods, average and maximum: 
\begin{align*}    
S_{\mathrm{avg}}(i)=\frac{1}{|\mathcal{H}(i)|}\sum_{h\in\mathcal{H}(i)} s_h(i),\qquad S_{\max}(i)&=\max_{h\in\mathcal{H}(i)} s_h(i).
\end{align*}
We compare the results in Table~\ref{tab: head-ablation}. For \textbf{Random}, simply adding more heads without meaningful partitions only decreases the amount of data to train each head, causing it to underperform the baseline. On the other hand, \textbf{All} improves modestly over LT/ST$_1$, showing that simply having more heads can indeed be beneficial, but it underperforms the human priors, LT/ST$_8$ or Item.

As for fusion, taking the average across heads might seem more natural, as commonly done in an ensemble. On the other hand, taking the maximum across heads might lead to overly optimistic estimates, similar to the issue of value overestimation in off-policy reinforcement learning \cite{KumarConservativeQLearning2020}. In fact, when we take the average on the random/all prior variant, or just using Temporal Prior, it does perform better than taking the max. However, we found that taking the max yields better results for Item Prior, in terms of both NDCG/recall and diversity. The reason might be each head evaluates whether the item candidate is a good fit under different criteria (e.g., item category or action type), as are done in multi-interest networks \cite{Li2019MIND, cen2020controllable}, while heads in random/all or Temporal Prior evaluate on more similar criteria. As future work, we may consider aggregating the scores hierarchically according to the nature of the prior type, and using inverse variance weighting to upweight more certain heads.

\begin{table}[H]
\centering
\caption{Head assignment vs.\ fusion on \textsc{Pixel8M}. Baseline here is LT/ST$_1$.}
\small
\setlength{\tabcolsep}{5.5pt}
\begin{tabular}{llcccccc}
\toprule
& & \multicolumn{2}{c}{Recall (\%)} & \multicolumn{2}{c}{NDCG (\%)} & \multicolumn{2}{c}{Entropy ($H$)} \\
\cmidrule(lr){3-4} \cmidrule(lr){5-6} \cmidrule(lr){7-8}
Prior & Fusion & {@10} & {@50} & {@10} & {@50} & {@10} & {@50} \\
\midrule
\rowcolor{lightgray} Baseline & ---  & 1.655 & 4.534 & 1.726 & 2.977 & 2.303 & 2.461 \\
\midrule
\multirow{2}{*}{Random}
& Max & 1.586 & 4.377 & 1.648 & 2.861 & \textbf{2.308} & 2.462 \\
& Avg & 1.602 & 4.417 & 1.670 & 2.893& 2.307 & \textbf{2.463} \\
\multirow{2}{*}{All}
& Max & 1.671 & 4.559 & 1.741 & 2.996 & 2.296 & 2.457 \\
& Avg & \textbf{1.672} & \textbf{4.560} & \textbf{1.743} & \textbf{2.998} & 2.301 & 2.460 \\
\midrule
\multirow{2}{*}{LT/ST$_{8}$}
& Max & 1.878 & 5.251 & 1.937 & 3.403 & 2.296 & 2.455 \\
& Avg & \textbf{1.950} & \textbf{5.338} & \textbf{2.028} & \textbf{3.501} & \textbf{2.303} & \textbf{2.457} \\
\midrule
\multirow{2}{*}{Item}
& Max & \textbf{1.749} & \textbf{4.695} & \textbf{1.834} & \textbf{3.115} & \textbf{2.371} & \textbf{2.515} \\
& Avg & 1.738 & 4.663 & 1.825 & 3.097 & 2.335 & 2.489 \\
\bottomrule
\end{tabular}
\label{tab: head-ablation}
\end{table}

\subsubsection{Long-Term vs. Short-Term Interests}
We study the generalization ability of LT/ST interests by training with $\tau=4$, and evaluating on $\tau\in\{1,4,8\}$ (Table~\ref{tab: long-short-term-interest-hllm}). For brevity, we only report the average gain over the baseline (next target prediction) across the eight metrics, and leave the complete results to Table~\ref{tab: hllm-structure-entropy}. All variants clearly surpass the baseline at $\tau=4$, and the benefit grows when we evaluate at $\tau=8$. On the other hand, when $\tau=1$, we only evaluate on the next one item as the target item, which is the same training objective as the baseline and only focuses on the short-term interest. Even under such a setting, also modeling the long-term interest leads to only a small drop in performance, which diminishes with more segments and disappears once we add the item prior. Finally, adding the item prior brings consistent benefit over the LT/ST prior alone, suggesting different human priors can be complementary.

\begin{table}[H]
\centering
\small
\setlength{\tabcolsep}{6pt}
\begin{tabular}{lccc}
\toprule
Variant & $\tau=1$ & $\tau=4$ & $\tau=8$\\
\midrule
LT/ST$_1$             & -3.07\% & 6.95\% & 7.72\%\\
LT/ST$_1$ + Item   & -1.23\% & 7.95\% & 8.34\%\\
LT/ST$_2$        & -2.68\% & 5.74\% & 6.71\%\\
LT/ST$_2$ + Item   & 0.25\% & 9.72\% & 10.36\%\\
LT/ST$_4$ + Item   & 0.04\% & 8.58\% & 9.07\%\\
\bottomrule
\end{tabular}
\caption{Average gain over the baseline HLLM (w/o prior) across eight metrics. We train the model on $\tau=4$, and show its robustness when evaluated on $\tau=1,4,8$.}
\label{tab: long-short-term-interest-hllm}
\end{table}

\begin{figure}[H]
    \centering
    \includegraphics[scale=0.3]{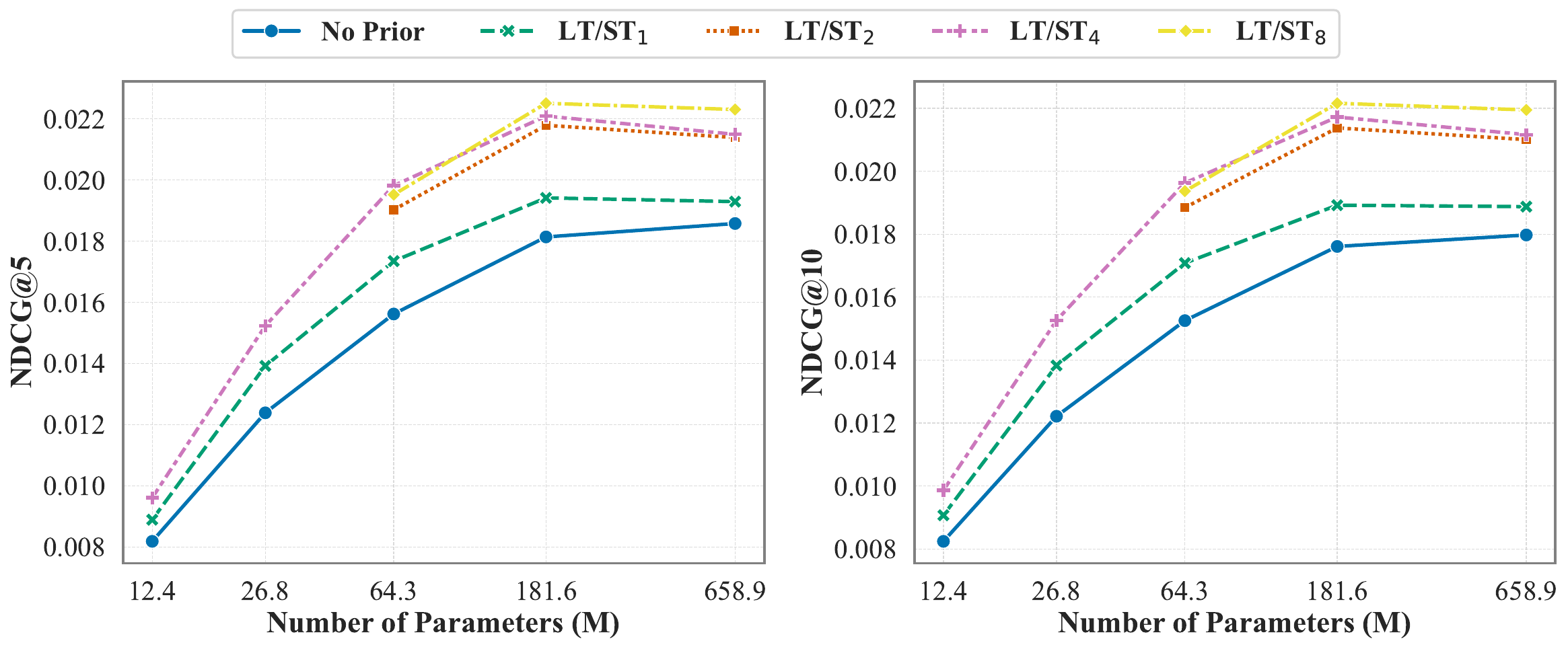}
    \caption{Sensitivity to the number of segments in Temporal Prior across different HSTU sizes.}
    \label{fig: long-short-term-interest-hstu}
\end{figure}

Next, we show that the benefit of LT/ST interests is consistent across different model sizes of HSTU (Figure~\ref{fig: long-short-term-interest-hstu}). Here, we train on $\tau=8$ using only the LT/ST prior. The baseline is only trained to predict the next one item, while LT/ST$_1$ means that we only model one interest, and the objective reduces to the simple discounted loss over the next $\tau$ items. We see that explicitly assigning more different heads to model the long-term vs. short-term interests brings more improvements as we go from the baseline to $\text{seg}=2$, and then the benefit from further increasing the number of segments slowly diminishes. We observe a slight dip from \texttt{size4} to \texttt{1b}, and our hypothesis is that the number of target items used to train each head decreases when the number of interests increases. Nevertheless, the benefit of separately modeling the long-term and short-term interests is still significant at the scale of \texttt{1b}, where the performance gain of the baseline model from increasing the model size has already flattened out at this scale.

\begin{table}[H]
\centering
\caption{HLLM results on Pixel8M with evaluation horizon $\mathbf{\tau=8}$. With multiple human priors, enforcing a \textit{hierarchical} composition (\textit{Hier}) yields the best Recall (R@K) and NDCG (N@K) while maintaining diversity (\textit{H}@K). A \textit{multiplicative} composition (\textit{Mult}) maximizes \textit{H} but underperforms on Recall/NDCG as it lacks explicit structure.}

\label{tab: hllm-structure-entropy}
\begin{tabular}{lrrr}
\toprule
HLLM Variants &   R@10 &  N@10 &  $H$@10 \\
\midrule
No Prior & 1.358 &    1.422 & 2.124 \\
LT/ST$_{1}$ & 1.464 & 1.523 &    2.126 \\
LT/ST${_1}$ + Item & 1.479 &    1.543 &    2.193 \\
LT/ST$_{2}$ & 1.442 &    1.496 &      2.121 \\
LT/ST$_{2}$ + Item (Mult) & 1.441 &    1.496 &      \textbf{2.223}  \\
LT/ST$_{2}$ + Item (Add) & 1.467&    1.524 &      2.193 \\
\midrule
\midrule
LT/ST$_{2}$ + Item (Hier) & \textbf{1.499} & \textbf{1.559} & 2.208 \\
\bottomrule
\end{tabular}
\end{table}

\subsubsection{Training Objective}
\label{sec: training-objective-ablation}
We conduct ablation studies on three key components of our training objective highlighted in Section~\ref{sec: train-obj}: 1) in-group negative sampling, 2) frequency balancing, and 3) the temporal discount factor $\gamma$. The results are summarized in Table~\ref{tab: weight-ablation}.

\begin{table}[H]
  \centering
  \caption{Ablation studies on the training objective. We report Recall (R@10), NDCG (N@10), and Entropy ($H$@10). The best results within each section are highlighted in bold.}
  \label{tab: weight-ablation}
  \begin{tabular}{lrrr}
    \toprule
    Ablation & R@10 & N@10 & $H$@10 \\
    \midrule
    Full Model ($\gamma=0.99$) & \textbf{2.000} & \textbf{2.099} & 2.3728 \\
    w/o in-group negative sampling & 1.642 & 1.700 & 2.6395 \\
    w/o frequency balancing & 1.928 & 2.005 & 2.4234 \\
    \midrule
    \multicolumn{4}{l}{\textit{Discount Factor $\gamma$}} \\
    $\gamma=1.0$ & 2.000 & 2.084 & 2.3773 \\
    $\gamma=0.95$ & 2.033 & 2.113 & 2.3768 \\
    $\gamma=0.9$ & 2.045 & 2.130 & 2.3727 \\
    $\gamma=0.8$ & 2.062 & 2.143 & 2.3768 \\
    $\gamma=0.7$ & \textbf{2.065} & \textbf{2.155} & 2.3773 \\
    $\gamma=0.6$ & 2.025 & 2.110 & 2.3810 \\
    $\gamma=0.5$ & 2.017 & 2.098 & 2.3818 \\
    \bottomrule
  \end{tabular}
\end{table}

First, removing in-group negative sampling (i.e., sampling negatives from all prior groups instead of the same group) and removing frequency balancing both lead to a sharp decline in Recall and NDCG. Specifically, removing in-group sampling causes R@10 to drop significantly from 2.000 to 1.642. This validates the importance of these mechanisms for learning effective representations and handling data imbalance.

The impact of the temporal discount factor $\gamma$ is more nuanced. Both Recall and NDCG peak at $\gamma=0.7$. Interestingly, the entropy $H$ initially decreases as $\gamma$ decreases from $1.0$ to $0.9$, while Recall/NDCG increases. However, beyond $\gamma=0.9$, the entropy $H$ instead increases alongside Recall/NDCG until they reach their peak at $\gamma=0.7$. This observation is intriguing because it suggests that, even for the same architecture, higher accuracy (Recall/NDCG) does not necessarily imply lower diversity (Entropy). Furthermore, the optimal $\gamma=0.7$ is notably different from the values typically used in reinforcement learning, which range from $0.9$ to $0.995$.

%% file: Sections/Appendix-Experiment-Details.tex
\section{Experiment Details}
\subsection{Dataset Details and Preprocessing}
\label{app:datasets}
\begin{table}[h]
\caption{Statistics of the datasets after preprocessing. Filtering criteria vary slightly depending on the backbone model.}
\label{tab:dataset-stats}
\centering
\small
\begin{tabular}{l l r r r}
\toprule
\textbf{Dataset} & \textbf{Backbone} & \textbf{\# Users} & \textbf{\# Items} & \textbf{\# Interactions} \\
\midrule
\multirow{2}{*}{Pixel8M} & HSTU & 561,737 & 398,261 & 57M \\
& HLLM & 2,220,506 & 404,182 & 102M \\
\midrule
MerRec & Both & 119,754 & 1,255,665 & 521M \\
\midrule
EB-NeRD & Both & 44,968 & 25,216 & 30M \\
\bottomrule
\end{tabular}
\end{table}

\textbf{Pixel8M.} For HSTU, we filter out users with fewer than 50 interactions. For HLLM, we filter out users with fewer than 20 interactions.
\begin{itemize}
\item Settings: For HSTU, we use a context length ($T$) of 50 and a forecast horizon ($\tau$) of 8. For HLLM, we use $T=10$ and $\tau=4$ due to its higher computation cost. 
\item Modalities: Text inputs are ``title'', ``tag'', and ``description''. Images are rescaled to $224 \times 224$ pixels.
\item Item Prior Details: Over 5\% of all items are labeled with the \textit{Miscellaneous} tag, which appear to be unlabeled data that can randomly come from any of the other tags. We have to simply set all 8 binary features to be True for this tag. However, this also demonstrates the robustness of our method to noise in the human prior. The resulting normalized frequency distribution of the eight features is: \textit{Entertainment}: 24.95\%, \textit{Real life}: 21.10\%, \textit{Performance \& Arts}: 15.30\%, \textit{Informational \& Educational}: 12.69\%, \textit{Fictional character}: 9.00\%, \textit{Music}: 6.29\%, \textit{Gaming}: 6.89\%, and \textit{Science \& technology}: 3.77\%. For example, the original tag ``Food Production'' is labeled with \textit{Real life}, \textit{Informational \& Educational}, and \textit{Entertainment}, while the original tag ``Celebrities Mix'' is labeled with \textit{Real life}, \textit{Entertainment}, \textit{Performance \& Arts}.
\end{itemize}

\textbf{MerRec.} We select the user subset with over 2,000 interactions.
\begin{itemize}
\item Settings: For HSTU, we use a context length of $T=400$, and for HLLM, we use $T=50$, both to predict one item ahead.
\item Modalities: Text inputs are ``c2\_name'' and ``brand\_name''.
\end{itemize}

\textbf{EB-NeRD.} We select the 44,968 users with over 512 interactions.
\begin{itemize}
\item Settings: For HSTU, we use $T=50, \tau=8$. For HLLM, we use $T=24, \tau=4$.
\item Modalities: Text inputs are ``title'', ``subtitle'', and ``topics''.
\item Graph Construction Details: For the co-engagement graph, we cap the contribution of highly active users by considering only their last 1000 interactions to prevent them from dominating the graph structure. When running the Leiden algorithm \cite{Traag2018FromLT} (implemented via the \texttt{igraph} package), we use modularity as the optimization objective and tune the resolution parameter based on the desired number of clusters. Small clusters falling below a size threshold are merged into a single larger cluster. 
\end{itemize}

\subsection{Model Architectures and Hyperparameters}
\label{app:models}

For both HSTU and HLLM backbones, we use a discount factor $\gamma=0.99$. However, we observed that a smaller $\gamma$ might lead to higher accuracy at the cost of lower diversity (Table~\ref{tab: weight-ablation}).

\textbf{HLLM Configurations.} The choice of LLMs for HLLM depends on the available modalities in the dataset. For Pixel8M, which contains both text and images, we use Qwen2-VL-2B-Instruct as the Item LLM and Qwen2.5-1.5B as the User LLM. For MerRec and EB-NeRD, which only contain text, we use TinyLlama-1.1B-Chat-v1.0 as the Item LLM and TinyLlama\_v1.1 as the User LLM.

\textbf{HSTU Configurations.} We experiment with various sizes\footnote{For the \textit{1B} model, we use the same hyperparameters as TinyLlama as in the original HLLM paper. However, the total number of parameters is less than 1B due to the simplification of the feed-forward blocks in HSTU.} of the HSTU model, detailed in Table~\ref{tab: model-sizes}. For all datasets, we remap product IDs based on the ones that still have been interacted with after filtering. We report the results based on size 3 unless noted.

\begin{table}[H]
\centering
\begin{tabular}{lrrrrrr}
\toprule
                & Size 1 & Size 2 & Size 3  & Size 4  & 1B  \\
\midrule
num\_layers     &      4 &      8 &      12  &      16  &   22   \\
num\_heads      &      4 &      8 &     8  &     16  &     32   \\
d\_model        &    128 &    256 &    512  &   1024  &    2048    \\
dropout      &      0.1 &      0.1 &     0.2  &     0.2  &     0.4   \\
\midrule
\midrule
Total Params  &  12.4 M &  26.8 M & 64.3 M & 181.6 M & 658.9 M\\
\bottomrule\\
\end{tabular}

\caption{The list of hyperparameters in the five model sizes of HSTU, along with the total parameter counts.}
\label{tab: model-sizes}
\end{table}

\textbf{Baseline Configurations.} For ComiRec and REMI, we use the self-attention version, which demonstrates performance comparable to the dynamic routing version, but with stabler and faster training. We use a learning rate of $0.001$, search the number of interest from $\{1, 2, 4, 8, 16, 32\}$, and report the best performance. For REMI, we tune the additional $\beta$ parameter for interest-aware hard negative mining from $\{0.1, 1, 4, 10\}$ and set routing regularization weight $\lambda$ to $100$ following the original paper.

For DualVAE, we reimplement it under the window-wise sequential recommendation setting. We search the aspect number from $\{4, 5, 10\}$ and set the VAE latent dimension to $32$. We search dropout from $\{0.1, 0.15, 0.2\}$, $\gamma$ (for contrastive loss) and $\beta$ (for KL-divergence) from $\{1e-3, 1e-2, 1e-1, 1\}$.

%% file: Sections/Appendix-Additional-Results.tex
\section{Case Studies}\label{sec: case-study}
\begin{figure}[ht]
    \centering
    \includegraphics[scale=0.8]{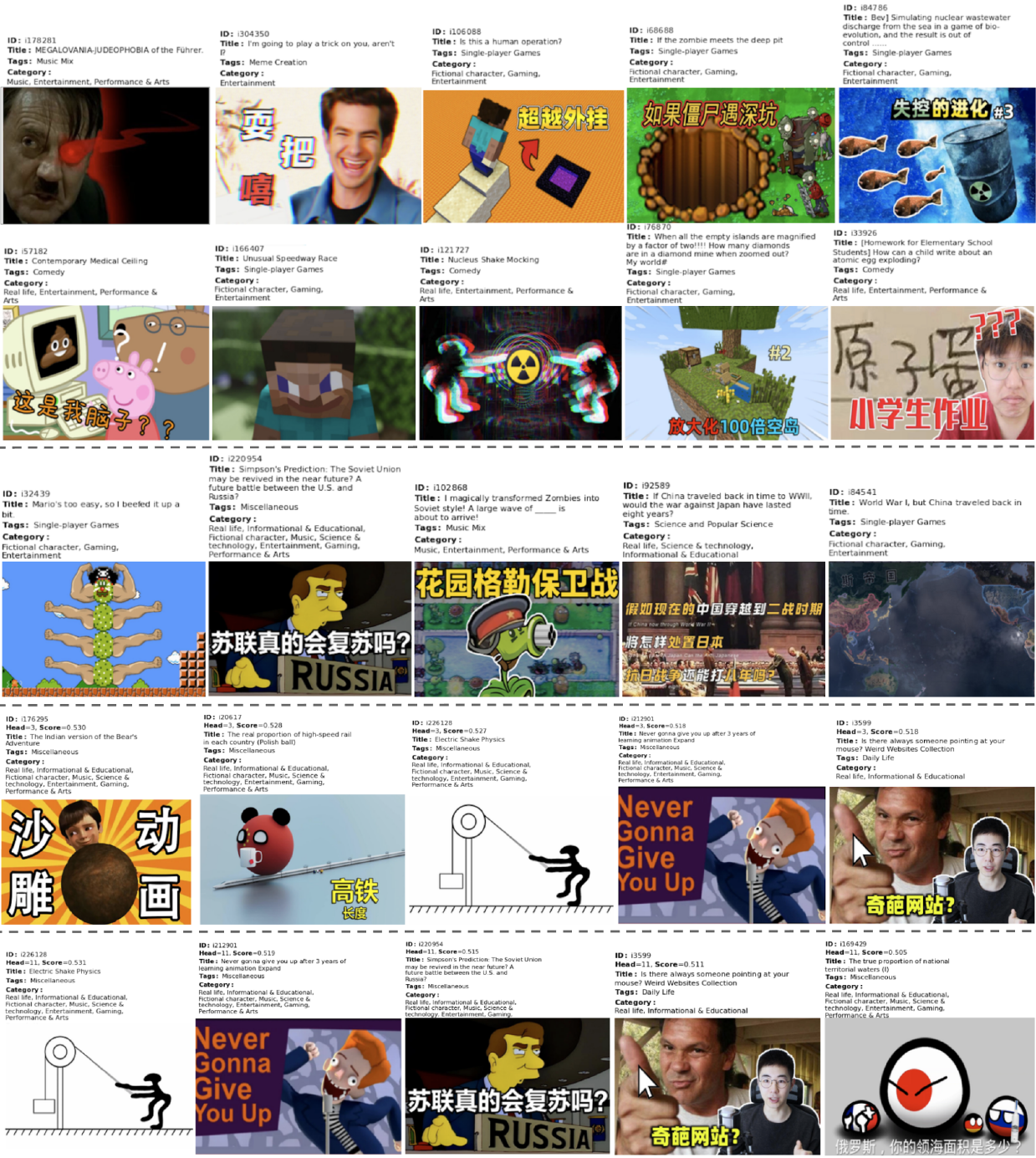}
    \caption{A case study on Pixel8M. Top two rows: browsing history. Third row: ground truth target items. Fourth row: HLLM baseline predictions (filtered to ``Informational \& Educational''). Last row: predictions from our model's long-term ``Informational \& Educational'' head.}
    \label{fig: case-study}
\end{figure}

To illustrate the qualitative benefits of human priors, we present a case study in Figure~\ref{fig: case-study} using the Pixel8M dataset. We analyze the predictions of the HLLM backbone, which leverages both text and image modalities\footnote{Video descriptions are omitted in Figure~\ref{fig: case-study} due to space constraints.}, configured with LT/ST$_2$ and Item Prior, and we compare it against the baseline HLLM.

The user history (top two rows) does not contain videos related to ``Informational \& Educational'', but they indicate a latent user interest in military and geopolitics, since they include clips from the WWII movie Downfall, a game on evolution under nuclear waste, and a meme on ``atomic egg explosion''. 

The baseline HLLM (fourth row) struggles to synthesize this nuanced intent. Even after filtered to the ``Informational \& Educational'' category, it produces relatively generic recommendations. In contrast, our proposed model demonstrates effective specialization (last row). The adapter head dedicated to the \emph{Long-Term} interest within the ``Informational \& Educational'' prior successfully captures the user's latent interest, resulting in recommendations such as the revival of the Soviet Union and Russian territorial waters. The better recommendation quality is validated by the target items (third row). Starting from the second target item, the user engaged with content related to the Soviet Union, which was successfully predicted by our model, and the Second Sino-Japanese War. This shows that dedicating a specialized head to model underrepresented categories is effective at discovering the user's hidden interests, which the baseline overlooked.

